\documentclass{article}

\begin{document}

\title{Emergent measure-dependent probabilities from modified quantum dynamics
without state-vector reduction}         
\author{M. B. Weissman\\
Department of Physics\\
University of Illinois at Urbana-Champaign\\
1110 West Green Street\\
Urbana, IL 61801-3080}
\date{}          
\maketitle

Counting outcomes is the obvious algorithm for generating probabilities in
quantum mechanics without state-vector reduction (i.e. many-worlds). This
procedure has usually been rejected because for purely linear dynamics it
gives results in disagreement with experiment. Here it is shown that if
non-linear decoherence effects (previously proposed by other authors) are
combined with an exponential time dependence of the scale for the
non-linear effects, the correct measure-dependent probabilities can emerge
via outcome counting, without the addition of any stochastic fields or
metaphysical hypotheses.
\section{Introduction}
The central question in the interpretation of quantum mechanics is how
unambiguous macroscopic observations arise probabilistically from an
underlying theory whose dynamical equation, in so far as it is known, obeys
superposition and is deterministic. (See, e.g. [1].) In this paper I shall
argue that the correct quantum probabilities may be obtained by
outcome-counting  [2, 3], the process which directly corresponds to our
operational definition of probabilities, in a model without state-vector
reduction, i.e. a type of many-worlds model.  To obtain this result, it
will be necessary to employ a modified state-vector dynamics, violating
superposition. I shall illustrate the idea using a cartoon form of
non-linear decoherence effects (previously proposed to arise in quantum
gravity [4, 5]), along with a simplified non-random form of the effectively
non-unitary time-dependence employed in recent explicit stochastic collapse
theories [6, 7, 8, 9, 10]. However, it will be unnecessary to invoke the
intrinsically stochastic constituents or metaphysical interpretive addenda
required in other approaches.

I shall begin by briefly reviewing related approaches to clarify what
problem this new approach is addressing. Then I shall present a toy model
to illustrate how the standard quantum probabilities can emerge as the
limits of ratios of numbers of distinct non-interfering outcomes in a
non-linear many-worlds picture, when non-linear decoherence processes
produce decoherent branches whose measures become equal in the long-time
limit. Some constraints on more complete and realistic models based on the
same idea will then be presented.

\section{Background: Probability with and without collapse}
Many-worlds interpretations have pointed out that the branching of the
state-vector into parts which represent macroscopically distinct outcomes
arises directly without alteration of the standard linear time-dependence.
[11, 12] Branching here means simply that the state evolves to one that can
be written as the sum of components between which so many variables (e.g.
particle coordinates) differ in complicated ways that the chance of
subsequent interference between the components is negligible. When parts of
a system become correlated with variables intrinsically outside the system
we can refer to the branching as true decoherence, for which subsequent
interference between branches is essentially undetectable.

The ``consistent histories" program of explaining the states along which
the branching preferentially occurs, within a linear dynamical equation
lacking an ``outside", [13] is far from complete. [14] In particular, if
the time-dependence of the state is purely linear, the time dependences of
the state components are entirely independent for {\em any} decomposition
of the state into components. Reference to interference {\em assumes}
physical significance of some non-linear function of the state (e.g. its
square), a feature which is not present in the linear dynamics.

 An open environment, however, can break the symmetry between states in a
system's Hilbert space, giving a natural set of ``pointer states" by a
process dubbed ``einselection"[15]. These pointer states are distinguished
by having a collection of predictable quasi-classical observables.[15]
Thus, despite the difficulties discussed above, it is possible to at least
consider defining probabilities of some preferred outcomes within a linear
theory without state-vector collapse.

It was realized long ago that, if one accepts the claim that
state-components representing macroscopically distinct outcomes form
non-interfering distinct ``worlds", the many-worlds interpretation leads to
an obvious prediction for the probabilities in simple experiments with a
finite number of possible discrete outcomes. [2, 3]  Since each world is
equally real, so is the version of the experimenter's mind which is
represented in that world and correlated with the other macroscopic
outcomes represented in that world. In an experiment repeated many times,
most worlds would show nearly equal numbers of each discrete outcome,
regardless of the measures of the state components representing those
outcomes. As Graham put it, ``It is extremely difficult to see what
significance measure can have when its implications are completely
contradicted by a simple count of the worlds involved, worlds that
Everett's own work assures us must be on the same footing." [3] The
prediction that probabilities of each macroscopically distinct outcome of a
quantum experiment would be equal, regardless of the measure of the
components of the state-vector, obviously contradicts observation. [2, 3]

Thus if the standard many-worlds (no collapse) view can explain
probabilities, it explains the wrong probabilities. As reviewed recently by
Saunders, [16] attempts to fix this probability problem within the
many-worlds interpretation of the linear time-dependence (e.g. [17, 18])
have required adding some metaphysical hypothesis beyond the evolving
quantum state, thus losing the initial appeal of deriving the
interpretation directly from the dynamical equation.

Recently, Zurek [15] has given a formal argument concerning the relation
between outcome-counting and probabilities. In order to extend the argument
to cases in which different outcomes have different measures, he makes a
formal decomposition of the density matrix into parts which have
correlations with various states of a hypothetical outside system. (See his
equations 4.10 through 4.12.[15]) The components of the density matrix
correlated with each individual outside state are implicitly assumed to
have equal measure, and a limit is considered in which the number of
outside states is so large that the number matched with any pointer state
of the system becomes proportional to the measure along that pointer state.

Zurek's procedure may be illustrated by considering a case in which only
two pointer outcomes (A and B) are possible for the system in state $\phi$.
The density matrix from Zurek's account is then:

\begin{equation}
\rho\ =\ \frac{\sum_{j=1}^a\ \mid A, j\rangle\langle A,j\mid\ +\
\sum_{j=a+1}^{a+b}\mid B,j\rangle\langle B,j\mid}{a+b}
\end{equation}

where the j's denote orthogonal states of the outside system and it is
postulated that $a/b\approx \mid <\phi\mid A>^2/<\phi\mid B>^2\mid$.
However, Zurek treats this decomposition as a purely formal procedure,
within the context of purely linear dynamical equations. If the states ``j"
and the density matrix (1) are fictive, it is difficult to see how they can
determine the operationally defined probabilities. If they represent a real
outcome of dynamics starting with a system state uncorrelated with the
outside, it is certain that those dynamics are not linear, since
superposition forbids the dependence of $a\ $ and $b\ $ on $<\phi\mid A>^2$
and $<\phi\mid B>^2$. Linear dynamics, of course, instead produce a sum of
{\em fixed numbers} of terms whose {\em coefficients} depend on $<\phi\mid
A>^2$ and $<\phi\mid B>^2$.

In contrast, macro-realist theories such as that of Ghirardi, Rimini, and
Weber (GRW) [19] or the more developed versions of Pearle and coworkers,
e.g. [7, 8], propose that there are explicit modifications of the dynamics,
which always directly or indirectly require non-linear time-dependence,
causing the state to collapse stochastically along some pointer projection
operator, rather than forming a superposition of macroscopically distinct
outcomes. [7, 8, 19] There is no time-reversed analog of the collapse, so
the dynamics do not obey CPT. The postulated set of pointer operators
directly addresses the preferred-state question raised by linear theories.
Whether these pointer operators must be introduced ad hoc or can in some
way be deduced from a deeper theory, perhaps including quantum gravity [4,
5, 8], remains to be seen. The key testable feature of such explicit
collapse theories is that the collapse process would cause anomalous
decoherence, so interference would be lost which would have been present if
there were strictly linear time dependence of the state. [20]

One argument supporting the existence of such non-linear processes is based
on the possibility that there are intrinsically unobservable variables (due
to horizons) associated with quantum gravity, so that the correlations
among the usual variables could only be represented by density matrix given
by a trace over the unobserved variables. [4, 5]  Within such genuine
decoherence theories, in general, the time derivative of the state-vector
is not representable by a linear operator and CPT symmetry need not be
obeyed. [4, 5]

The better-developed versions of the explicit collapse theories require the
inclusion of a non-quantized stochastic field. [7, 8, 9, 10]  These
stochastic fields possess non-local correlations to account for the results
of Bohm-EPR-Bell experiments. [21] An attempt to quantize the stochastic
field produced only decoherent superpositions, not actual collapse. [6, 10]
A ``moving finger" was then hypothesized to pick which decoherent solution
actually occurred, [10] similar to the hypotheses outside the dynamical
equations required in other approaches. Thus, if the explicit collapse
theories are to dispense with metaphysics, they appear to require not only
a non-linear dynamical equation but also explicit random elements and
classical fields, leaving a theory with very diverse constituents.

\section{A new approach}

The idea I will present essentially amounts to a justification for how
density matrices resembling that in equation (1) can arise within
non-linear dynamics, thus justifying their use in predicting actual
probabilities. In other words, the incorrect probabilities which have been
predicted to arise from counting branches in the linear dynamical theory
may be fixed by altering the dynamical equations rather than by giving up
the compelling outcome-counting algorithm. Although I shall borrow the
ideas of  a non-linear loss of interference and an intrinsic time asymmetry
from arguments employed in explicit collapse pictures, [4, 5, 6, 10, 19],
by dropping the requirement of unique outcomes  I shall remove the need for
non-quantized stochastic fields. (Many-worlds interpretations have always
pointed out that the consistency of observed macroscopic reality does not
require macroscopic outcomes to be unique unless one makes the auxiliary
assumption that the observer remains unique.[11])

The essential idea is that, as found in the attempt to fully quantize
collapse theory by replacing the classical stochastic field with a quantum
field representing new variables, [6, 10] each macroscopic branch of the
state-vector has many sub-branches, with distinct values of some new
quantum variables. However, rather than arbitrarily consign all but one of
these sub-branches to ``non-reality", we assume that they all actually
persist. Thus probabilities are determined simply by their numbers, as in
the obvious outcome-counting algorithm which gave incorrect probabilities
in purely linear collapse-free theories. In order for the newapproach to
work, the number of sub-branches associated with each macroscopic outcome
must become proportional to the measure of the component of the state
representing that outcome on the time-scale of a measurement, i.e. on the
same time-scales invoked in explicit collapse theories, for the sorts of
macroscopic systems in which quantum probabilities have been recorded. [9,
19]

In the discussion to follow, ``sub-branch" has a fairly precise meaning: a
state component which cannot be divided into distinct parts between which
subsequent interference is vanishingly small, when all the quantum
variables are included. A ``sub-branch" would be called a ``world" in
many-worlds terminology. (It will turn out not to be necessary to specify
ahead of time whether two macroscopically distinct outcomes are always on
distinct sub-branches, since if sub-branches containing macroscopic
superpositions are allowed, they will become an exponentially rare fraction
of the total number of sub-branches.) ``Branch" will have a less precise
meaning: a collection of sub-branches which comprise a macroscopically
distinct outcome. When a continuum of macroscopic outcomes are possible,
this grouping is a bit arbitrary, as usual, but none of our arguments will
rely on how that grouping is made.

\section{Illustration via a toy model}

We may illustrate the basic workings of approach by using an initial toy
model, closely analogous of GRW. [19] This toy model is not relativistic,
and is not intended to give an accurate representation of the non-linear
branching dynamics. Instead, I simply assume a particularly simple
non-linear branching algorithm to illustrate the generic consequences of
such processes, whose origins and details must be sought elsewhere.

We assume that the Hilbert space in which the statevector $\mid\Phi >$
resides is the direct product of the space in which the conventional state
$\mid\phi >$ resides and a space which describes new variables. The state
of these new variables will be denoted by a label L, whose form will be
discussed below. Then, a basis set for the whole space can be formed from
states of the form $\mid\Phi >=\mid\phi>\mid L>$. We assume that there is
set of pointer projection operators $\{P_\Gamma\}$, where each $\Gamma$
denotes both a sub-space (denoted $\gamma$) of the standard Hilbert space
and a particular sub-space of the new-variable space. These pointer
operators will not be be assumed to be normalized, i.e. $P_\Gamma^2 =
pP_\Gamma$ with $p\not= 1$ in general. In fact (for reasons soon to be
explained) they will be assumed to have a time-dependent normalization and thus
be written as $P_\Gamma(t)$.

For the non-linear branching process, we employ the simplest cartoon of
such a process we can come up with. When any $M_\Gamma (t)\equiv <\Phi\mid
P_\Gamma (t)\mid\Phi >$, reaches a definite value (which we can arbitrarily
define to be unity), a new sub-branch emerges, on which the unspecified new
variables acquire a new state orthogonal to their old state as described by
the following algorithm

\begin{eqnarray}
\mid\Phi \rangle \rightarrow \frac{Z^{1/2} \langle\Phi\mid P_\Gamma
\mid\Phi \rangle }{\langle\Phi \mid P^2_\Gamma \mid\Phi \rangle }\ \
C_\Gamma (t) P_\Gamma \mid\Phi \rangle  \ + \nonumber\\
\left\{ \mid\Phi \rangle \ -\ \frac{(1-(1-Z)^{1/2})\langle\Phi \mid
P_\Gamma \mid\Phi \rangle }{\langle\Phi\mid P^2_r \mid\Phi \rangle }\ \
P_\Gamma\mid\Phi \rangle \right\}
\end{eqnarray}

Here we have described the development of an orthogonal state of the new
quantum numbers on the new sub-branch by the unitary operator $C_\Gamma(t)$
which acts on the Hilbert space of the new variables. Z is a real number,
$0<Z<1$, which specifies the fraction of the measure in the $\Gamma$
sub-space  which is assigned to the new sub-branch. The expression in curly
brackets represents the old state with some of its measure in the $\Gamma$
pointer sub-space removed. The coefficient of the second term within the
brackets has been picked to keep the norm $<\Phi\mid\Phi >$ of the total
state unchanged. We shall refer to the dynamics represented by algorithm
(2) as anomalous branching.
 Algorithm (2) is written in a form independent of the normalization of
$P_\Gamma$ and also independent of the dimensionality of the subspace to
which $P_\Gamma$ projects. If two distinct $M_\Gamma$'s reach 1 at the same
time, it is easy to check that the state resulting from the bifurcation
algorithm is independent of the order in which the two branchings occur, so
long as the product of the two projection operators is zero. The parameter
Z in algorithm (2) unfortunately remains arbitrary in the absence of a
deeper theory. For certain illustrative purposes, we shall employ Z=1/2,
but shall argue that, although no fine-tuning of Z is required, the simple
choice Z=1/2 is not in general suitable. Most details of the splitting
process (even our choice of bifurcations rather than some more general
multifurcation) will not, however, be essential to the workings of the
model, so long as it preserves $<\Phi\mid\Phi>$.

Although I have nothing new to add to previous ideas concerning the origins
of non-linear decoherence, [4, 5] a cartoon account of the operator
$C_\Gamma (t)$ might be helpful. One could imagine the Hilbert space of the
new variables as something associated with the as yet unrealized quantum
theory of gravity and $C_\Gamma(t)$ as the creation operator for a particle
propagating endlessly into an open ``outside", starting at time t.
Permanent maintenance of orthogonality between states created at different
times might require either some non-linearity to avoid dispersion or
horizons to make dispersion irrelevant.

We assume that there is an initial unique state for all the new variables
(of which there may be an uncountable number) so all that needs to be
expressed to distinguish their current state is a countable list of the
changes in that initial state induced by the anomalous branching process.
One may then regard $C_\Gamma (t)$ as adding a label ($\gamma$,t)
specifying the pointer operator involved and the time at which the
anomalous branching occurred:

\begin{equation}
C_\Gamma (t)\mid \Phi_\Gamma \rangle\ \equiv\ C_\Gamma
(t)\mid\phi_\gamma\rangle\mid L_O\rangle\ =\ \mid\phi_\gamma\rangle\mid
(\gamma ,t),L_O\rangle
\end{equation}

where $L_O$ is a list of previously acquired labels. Each sub-branch here
acquires a unique list of labels recording the previous anomalous branching
times for each pointer operator. The labeling (although not the branching
algorithm) closely resembles ones employed both in linear decoherence
theories [15] and in non-linear collapse pictures. [9, 10]

Processes like algorithm (2) would not suffice to maintain non-linear
branching if each $P_\Gamma$  maintained a fixed normalization. The reason
is that once the measures of the separately labeled (permanently
decoherent) sub-branches fell to less than $1/p$ there would be no way for any $M_\Gamma$ to reach
one again. Therefore it is necessary, in order to obtain a theory in which
non-linear branching persists, to invoke a non-unitary time-dependence of
the pointer operators:

\begin{equation}
P_\Gamma (t)\ =\ P_\Gamma (0)e^{t/\tau}
\end{equation}

Although there will be one particular time (before the first anomalous
branching) for which $P_\Gamma (t)$ is a normalized (p=1) projection
operator, that need not be chosen as t=0. At this point we shall not
discuss how to assign $\tau$, except that the same $\tau$ will be used for
each pointer operator within any family related by a symmetry such as
spatial translation.

Let us initially consider a particle for which we can ignore the ordinary
quantum dynamics, e.g. because the particle is very massive. We shall also
assume that the projection operators are one-dimensional, i.e. they project
to states. Let us also initially assume for convenience that the parameter
Z=1/2.

We shall follow what becomes of a state component (with a single initial
label, $L_O$) whose state $\phi$ happens to have been split by  a
``measurement" into two widely separated pieces A and B, each conveniently
lying along one pointer state. In other words, there are pointer operators
$P_{A,L}\ =\\ ge^{t/\tau}\mid A>\mid L><L\mid <A\mid$ and $P_{B,L}\ =\
ge^{t/\tau }\mid B>\mid L><L\mid <B\mid$  where g is a constant whose value
depends exponentially on the choice of time origin.

We first consider the case in which $M_{A,Lo}(t)= M_{B,Lo}(t)$. For
convenience we assign t=0 to the time of the first branching, which is
equivalent to setting the prefactor g in $P_{A,L}$ to g=2 if we choose to
express this state component in normalized form. Applying algorithm (2)
gives simultaneous branching for A and B, with the resulting state
independent of the order in which we perform the two branching operations.
We define T$\equiv \tau ln(2)$, the time between successive paired
branching events. We obtain the following sequence of states  after these
events:

\begin{eqnarray}
& &1/2^{1/2})(\mid A\rangle + \mid B\rangle )\mid L_O\rangle\ \
\rightarrow\nonumber\\
& &(1/2) (\mid A\rangle\mid (A,0), L_O \rangle\ +\ (\mid A\rangle + \mid
B\rangle) \mid  L_O\rangle + \mid B\rangle\mid (B,0), L_0 \rangle )\ \
\rightarrow\\
& &(1/2^{3/2})(\mid A\rangle (\mid (A,T),(A,0),L_O \rangle + \mid (A,T),
L_O\rangle + \mid (A,O),L_O\rangle )\ +\nonumber\\
& &(\mid A \rangle + \mid B\rangle )\mid L_O \rangle + \mid B\rangle (\mid
(B,T), (B,O),L_O \rangle + \mid (B,T),L_O \rangle + \mid (B,O),L_O\rangle
))\nonumber
\end{eqnarray}

It is easy to check that after J doubling times, there will be $2^J-1$ each
of the distinctly labeled A and B sub-branches as well as a component of
the original form $(\mid A>+\mid B>)\mid L_o>$. Whether this latter
component is to be considered one or two sub-branches, i.e. whether
subsequent interference between $\mid A>$ and $\mid B>$ is possible,
becomes irrelevant as it becomes exponentially out-numbered by
macroscopically distinct sub-branches. Since the fundamental postulate is
that each sub-branch has equal probability, the probability of such a
macroscopic superposition falls to zero exponentially in time, while the
two distinct macroscopic outcomes remain equally likely.

Now let the initial state have 2/3 of its measure along A and 1/3 along B.
We choose g=1.5 to set t=0 as the time of the first branching, which occurs
only for A. Subsequent branching is again simultaneous for A and B. We
obtain the following sequence of normalized states, in which the notation
$L_0$ (shared by all these sub-branches) has been suppressed in all the new
branches, for brevity:

\begin{eqnarray}
&\ &(1/3^{1/2})(2^{1/2}\mid A\rangle\ +\ \mid B\rangle )\mid L_0 \rangle\ \
\rightarrow\nonumber\\
&\ &(1/3^{1/2})(\mid A \rangle\mid (A,0)\rangle\ +\ (\mid A\rangle\ +\ \mid
B\rangle )\mid L_0 \rangle )\ \ \rightarrow\\
&\ &(1/6^{1/2})(\mid A\rangle (\mid (A,T),(A,0)\rangle + \mid (A,T)\rangle
+ \mid (A,0)\rangle )\ +\ \nonumber\\
&\ & (\mid A\rangle + \mid B\rangle )\mid L_0\rangle + \mid B\rangle \mid
(B,T)\rangle )\nonumber
\end{eqnarray}

It is easy to check that after another (J-1) doublings, there will be
$(2^{J+1}-1)$ pure-A sub-branches with distinct labels, $(2^J-1)$ pure-B
sub-branches with distinct labels, and one state component of the form
$(\mid A>+\mid B>)\mid L_o>.$ If one considers this component as consisting
of one A and one B sub-branch, the ratio of the number of A to B
sub-branches becomes exactly two. Again, even if one does not {\em a
priori} rule out macroscopic superpositions as possible outcomes, the
probability of such a superposition falls to zero, as discussed above, and
the ratio of the numbers of A to B sub-branches approaches two, with the
difference from that limit decreasing exponentially in time.

The two examples above were contrived so that the branching along the two
pointer states occurred synchronously, by making the initial ratio of the
measures along those pointer states a power of two, matching the factor by
which the measure along a pointer state changes in the branching algorithm
with Z=1/2. For different initial ratios of the measures of those
components, the ratios of numbers of A andB sub-branches would oscillate as
first the A sub-branches, then the B sub-branches, bifurcated. The
definition of the time-averaged limiting ratio of the numbers of
sub-branches of each type would then be slightly arbitrary. That problem is
an artifact of the choice Z=1/2, which was employed only for illustrative
simplicity, not on the basis of any real dynamical theory. For a more
general bifurcating case, in which ln(Z)/ln(1-Z) is irrational,
synchronization becomes impossible since bifurcations can give rise to new
values of $M_\Gamma$'s, distinct from other sub-branches. In the more
general discussion which appears in a later section, I shall argue (on the
basis of simulations and of analytic arguments a bit short of a proof) for
a conjecture that in this case initial synchronization becomes unnecessary
because the distribution of $M_\Gamma$'s, of descendant sub-branches will
spread and acquire a branch-independent mean. Since the $M_\Gamma$'s, are
simply standard measures scaled by a common time-dependent factor, one
obtains a branch independent ratio of number of sub-branches on a branch to
the standard measure of the branch, with vanishing fluctuations around that
limiting ratio.

We next give an example of the emergence of standard probabilities when the
macroscopic branches are spread out over different numbers of pointer
states. For example, let us suppose that the pointer states $\phi_\gamma$
are minimum-uncertainty Gaussian states of a single particle spread over
regions with width {\em w}, in analogy with the GRW collapse picture. [19]
Now let piece A initially have 2/3 of the measure and have a Gaussian
spread with a width of 400{\em w}, and piece B have of 1/3 of the measure
with a width of 100{\em w}.  The pointer states here are uniformly
distributed in three-dimensional real-space, so (with the above choices of
widths) A is spread over 64 times as many pointer states, and hence (with
those choices of measure) has $2^{-5}$ as much measure on each pointer
state. Therefore its anomalous branching will begin later than the
anomalous branching of the B components by a time 5T. Thus at a particular
part of the Gaussian distribution (e.g. at the peak, or at the shell one
standard deviation from the peak) the components of B along each pointer
state will split into 32 sub-branches before the corresponding components
of A also start to branch. However, A is distributed along 64 times as many
pointer states. Therefore the total number of A sub-branches becomes just
twice the number of B sub-branches, as expected from their measures,
despite the asymmetry between their spatial distributions.

We have seen examples in which the standard probabilities emerge when
ordinary quantum rates of spreading between pointer states are negligible
compared to the anomalous branching rates. Except for the issue that the
ratio of the numbers of sub-branches on two branches can oscillate, the
basic procedure did not depend on special features of the initial states.
However, {\em those probabilities are not just built-in by fiat, but rather
arise by dynamics.}  In fact, immediately after a ``measurement" process,
i.e the development by ordinary quantum dynamics of a state with
substantial projections on distinct pointer sub-spaces,  non-standard
probabilities would appear transiently before the anomalous branching has
occurred many times.

Now we can consider the effect of the standard quantum dynamics on the
probabilities. The examples will show that the branching algorithm is not
guaranteed to produce standard quantum probabilities when applied to an
isolated microscopic system (for which the rate at which quantum dynamics
spreads the state out over different pointer sub-spaces is not negligible
compared to the anomalous branching rate), but that once enough particles
are coupled to form a macroscopic system, ordinary probabilities emerge.

Let us say that the state representing a particle with mass m, with
pointer-state growth-rate parameter $1/\tau_1$, has just split at t=0 into
(among others) a piece which initially lies along one of the minimum
uncertainty pointer states, with $M_\Gamma =1/2.$ The velocity spread
$\hbar$/(2{\em w}m) will give an increase with time of the number of
pointer states over which the resulting state is spread, with the maximum
measure along the initial state decreasing as a result. Then the maximum
$M_\Gamma$ will be  $M_\Gamma (t) = exp(t/\tau_1)(1+(t/t_0)^2)^{-3/2}/2,$
where $t_0 \approx w^2m/\hbar$. If $\tau_1 \gg t_0, M_\Gamma(t)$ will not
reach one again until $t=\tau_1(ln(2)+3ln(t/t_0))\approx 3\tau_1
ln(\tau_1/\tau_0)$. The normal dynamics then delay the time for the
anomalous branching by a factor of approximately $3ln(\tau_1/t_0)$. If we
use GRW parameters for a proton, i.e. rate $1/\tau_1 =10^{-16}s^{-1}$ and
width {\em w}$=10^{-5}$ cm we find $t_0 \approx 10^{-7} s,$ and the
logarithmic factor $3ln(\tau_1/t_0)$ would be about 150.

This example illustrates several points. First, a sufficiently small
exponential growth rate for single-particle pointer operators, like the
small GRW single-particle decoherence rate [19], allows the standard
quantum dynamics to proceed uninterrupted for a very long time. In fact, in
this model (unlike the GRW model, which has Poisson collapse statistics
[19]) the reduction of the sub-branch measure by a branching event actually
shuts down the anomalous process for a long time subsequently, so that the
branching events are separated by intervals longer than $\tau_1.$ However,
the spreading from the pointer states due to the uncertainty-principle (or
more generally, the non-commutivity of the pointer operators with the
Hamiltonian, H) means that each sub-branch spreads out over many pointer
states before the next anomalous branching. As a result, the anomalous
branching process would not fix the mean sub-branch measure independently
of the quantum dynamics, which can differ among branches.

In this single-particle example, the measure which starts on a single
pointer state after an anomalous branching spreads out over a very large
number, about $(\tau_1/t_0)^3\approx 10^{70},$ of pointer states before the
next round of anomalous branching starts. Therefore, if one were to assign
probabilities of the particle being in different locations by counting
sub-branches in this isolated microscopic system, those probabilities would
not be proportional to the square of the wave-function. For example, after
about half of this state has undergone another splitting, half the measure
would be represented by some $10^{70}$ sub-branches covering the spatial
middle part of the state, while the half of the measure contained in the
outer parts would be represented by only one sub-branch.

In the example above the deviations from the standard probabilities are
quite striking in an isolated  microscopic system, although any assignment
of probabilities would make little sense on a short time scale, for which
there is no stable ratio of numbers of sub-branches representing different
outcomes. As most discussions of quantum measurement routinely point out,
measurements can never be made on {\em isolated} microscopic systems,
because any measurement for which we can possibly have tabulated
probabilities requires that correlations be established with macroscopic
variables, at least including ones describing our brains. Thus the
requirement for a successful theory (assuming no special role for
consciousness) is that it predict the standard probabilities for a large
class of macroscopic systems, to which we now turn for further examples.

First, consider a particle of larger mass. The spreading time $t_0$ for the
simple Gaussian pointer states scales linearly with m. Thus, even if $\tau$
were m-independent, a more massive particle would have $\tau < t_0$ for $m
> 0.1 g= (\tau_1/t_0)10^{-24}g.$ Within the explicit collapse framework, it
has been noted that experimental constraints indicate that the anomalous
decoherence rate should increase with rest mass. [8, 9, 10] If, for
example, $1/\tau$ were to scale as m, the massive particle would have $\tau
< t_0$ for $m > 10^{-12} g.$

Thus for sufficiently massive particles, regardless of how the quantum
process had dispersed the wave-function, measure-proportional probabilities
would emerge. If 1/$1/\tau$ were proportional to m, branching times would fall to less than one microsecond for $m > 0.1 g.$

\section{Multi-particle systems}

Now we consider a system of N particles, which for simplicity we make
distinguishable. The model should allow branching to occur for any one
particle without affecting the others, so that for uncorrelated particles
the branching will not create spurious correlations. Therefore each pointer
operator should be a product of a pointer operator for one particle by
identity operators for the other N-1, e.g. $P_{1\Gamma}I_2 \ldots I_N.$
Each pointer operator then projects  not to a state of the overall system,
but instead to a higher-dimensional sub-space. The identity operators for
the other particles apply to both the standard Hilbert space and to the
space represented by the corresponding labels. We shall assume that these
pointer operators are also appropriate for correlated multi-particle
states. The formalism is close enough to that of GRW [19] to not require
repetition in detail.

Let us consider a crystal in its internal ground state, so that only the
center-of-mass coordinate remains free. Assume that the masses of the
(distinguishable) particles are all still m, and the pointer operators each
still have an exponential growth rate $1/\tau_1.$ The spreading time
$t_{0N}$ would be the same as for any other particle of mass Nm, i.e.
$Nt_0.$ Now as soon as any particle's state branches, the center of mass
coordinate of the crystal is localized to within distance {\em w} on each
new sub-branch, just as in the GRW theory. [19] After $N_B$ repeated
branching events most sub-branches are localized to within {\em
w}$/N_B^{1/2},$ as is the unique state in the analogous collapse process in
GRW.

The crucial question then becomes how long it takes for anomalous branching
to occur in this multi-particle collective state. {\em The key point is
that since the N particles arrived in the crystal through partially
independent histories, the timings of the cyclings of their $M_\Gamma$
between branching events are not the same. For long enough and complicated
enough histories, one expects these timings to be uncorrelated.} Therefore
the logarithms of those $M_\Gamma$ (on any one of the allowed product
states) are randomly distributed, e.g. from -ln(2) to 0 if we follow
algorithm (2) with Z=1/2 and if $Nt_0 >> \tau_1/N.$ Branching events can
occur at any time, with the typical time delay between branching events
reduced by a factor of N compared to the single-particle rate.

Here our procedure is different from GRW [19] in an important way (in
addition, of course, to the interpretation that only branching, not
collapse, is occurring). {\em The randomness in the timing of the branching
events is not put in as a separate fundamental ingredient here.}  Rather,
it arises from the complicated but deterministic histories of the
constituent particles. In other words, {\em it is an ordinary statistical
mechanical effect, not an essential randomness in the constituent dynamics.}

\section{More General Considerations}
I have illustrated how non-linear decoherence effects can give standard
quantum probabilities via outcome counting using a toy version of the
non-linear dynamics, applied to simple cases. Although I am unable to
develop a full theory of respectable non-linear dynamics, it is nonetheless
possible to discuss some of the constraints on such a theory and to clarify
why the non-linear procedure above produced correct probabilities in the
examples given.
The key points in the dynamics leading to the measure-proportional
probabilities are that:

\noindent I.\ \ The anomalous sub-branching does not change the net measure
on any branch.

\noindent II.\ \ A steady-state is approached in which the mean value of
the measures of the sub-branches of a given branch is independent of the
branch.

{\em If these two conditions are met, then the sub-branch numbers must be
proportional to branch measure.}

The condition (I) is assumed to apply exactly to the anomalous branching,
as in algorithm (2). The general form of the anomalous branching rate, a
monotonically increasing function of $M_\Gamma$, will tend to cause
condition (II) to be met approximately, but it will be met precisely only
under further constraints.

Let us first consider the case in which the ordinary quantum dynamics can
be ignored. The measures of the sub-branches of the different branches
became precisely equal at all times in our simplest (Z=1/2) examples only
because of the specially contrived initial measures. In general, to insure
that the average over equally weighted sub-branches of $M_\Gamma$ (denoted
$<M_\Gamma >_C)$ approach the same limiting values  on different branches,
one needs that $<M_\Gamma >_C$ approaches the mean of a limiting
steady-state distribution. I conjecture that that will occur whenever
ln(Z)/ln(1-Z) is irrational.

There is an informal argument for the conjecture above. The branching
algorithm together with the pointer-operator growth implies that after time
t, the log of the measure of each sub-branch will have to be reduced by
$t/\tau$ to within an accuracy of max $(\mid ln(Z)\mid, \mid ln(1-Z)\mid).$
The reduction occurs in steps of ln(Z) and ln(1-Z)  The number of ways that
$t/\tau$ can be put together (to the specified accuracy) out of
combinations of the form $(J_Zln(Z) +J_{1-Z}ln(1-Z))$ with positive integer
J's is of order $t/\tau.$ Each such combination leads to a different
sub-branch measure, a set of pseudo-random numbers. Although at finite time
the probability density function $\rho (M_\Gamma)$ of the set of
$M_\Gamma$'s is always a finite collection of delta functions, it seems
likely that its moments will approach those of a distribution which is
time-independent under the branching algorithm. (It is not too hard to
demonstrate that for rational ln(Z)/ln(1-Z)= m/n, in reduced rational form,
the distribution of the number of $M_\Gamma$'s in n uniformly spaced
logarithmic bins from ln(Z) to 0 approaches a unique limit, implying that
$<M_\Gamma >_C$ becomes fixed to within a factor of $Z^{1/n}$. Formally
taking a suitable limit to obtain the fixed-mean result for irrational
ln(Z)/ln(1-Z) does not seem to be trivial.)

The time-independent distribution for the bifurcation scheme can fairly
easily be shown to be:

\begin{eqnarray}
\rho(M_\Gamma )\ =\ \frac{Z'}{M_\Gamma^2}\ \ \ {\rm for}\  Z'\leq M_\Gamma
<1-Z' \nonumber\\
\rho(M_\Gamma ) = \frac{1}{M_\Gamma^2}\ \ \ {\rm for}\  1-Z'\leq M_\Gamma
\leq 0
\end{eqnarray}

where $Z'\equiv$ min(Z, 1-Z). For this limiting distribution $<M_\Gamma
>_C$ will be Zln(1/Z)+(1-Z)ln(1/(1-Z)). For some more general branching
scheme, so long as the moments approach those of any  well-defined
distribution, condition II will be met.

One can run simulations of the branching algorithm to see if the moments do
actually converge to the ones predicted from distribution (7). Simulations
with ln(Z)/ln(1-Z)= (1+5$^{1/2}$)/2 were run to times of t=8000$\tau$.
$<M_\Gamma >_C$ showed irregular fluctuations as a function of t, around
the limit  calculated from distribution (7). The envelope of the
fluctuations was a decreasing function of t, with $\mid\Delta ln<M_\Gamma
>_C\mid$ staying under 0.03 for 25$\tau < t < $8000$\tau$, under 0.015 for
150$\tau < t < $8000$\tau$, and under 0.003 for 4300$\tau < t <
$8000$\tau$. (Slightly quicker simulations of $\mid <\Delta ln M_\Gamma
>_C\mid,$ nearly identical to $\mid\Delta ln<M_\Gamma >_C\mid$ for $t
>10\tau$, extended these limits to 13500$\tau$.) The simulation results are
consistent with the plausible hypothesis that the deviations of $<M_\Gamma
>_C$ from its limiting mean scale as (t/$\tau)^{-1/2}.$

The key issue determining whether condition II (branch-independent
\newline $<M_\Gamma >_C$) is met is not the mathematical exercise required to show a
limiting $<M_\Gamma >_C$ in a given branching algorithm, but rather what
the effects are of including ordinary quantum dynamics. Given that events
happen, the pointer operators cannot commute with H, so that even if a
sub-branch starts out with a well-defined value of a pointer operator, it
will not keep it.

I have argued that a pure branching dynamics of the type described will
give a distribution of $M_\Gamma$ whose mean is independent of the initial
state. The mean of the sub-branch measures on each branch will approach
$e^{-(t/\tau)}<M_\Gamma >_C$ so long as the ordinary quantum processes have
little effect on the anomalous branching. That requires that a sub-branch
which starts off as an eigenstate of some $P_\Gamma$ spreads very little
outside the $\Gamma$ subspace before the next anomalous branching. Starting
in an eigenstate of $P_\Gamma$, the first derivative of the measure of the
projection of the state onto the $\Gamma$ subspace will be zero. The second
derivative will be negative and will depend on the commutator $[P_\Gamma
,H]$ of the pointer operator $P_\Gamma$ with H. We then require:

\begin{equation}
\tau_{eff}^2\ \ \frac{\langle\Phi_\Gamma \mid [ P_\Gamma
,H]H\mid\Phi_\Gamma\rangle}{\hbar^2\langle\Phi_\Gamma \mid
P_\Gamma\mid\Phi_\Gamma \rangle}\ \ \ \ \ll 1
\end{equation}

where 1/$\tau_{eff}$  is the effective anomalous branching rate for the
system. As discussed in the toy example, there is reason to expect that for
macroscopic objects 1/$\tau_{eff}$ would scale linearly with the size of
the system. [7, 19] An essentially similar condition was found by GRW [19]
as a requirement for the existence of classical trajectories in collapse
pictures with stochastic ingredients in the dynamics.  Although we are far
from having a suitable general choice of pointer operators, quantum
commutators of operators representing measurables, e.g. [P,H], generally
become relatively unimportant for large systems. It is then plausible that
condition (8) will be met by macroscopic systems in general.

\section{Remaining problems and prospects}

We have seen that in a toy model correct quantum behavior is retained for
small enough objects, while for large objects the probabilities obtained
from outcome-counting approach values proportional to quantum measure. The
essential features of that toy model seem likely to generalize.
Nevertheless, this proposal is obviously in an early stage, with some
elements directly borrowed from the GRW stochastic collapse proposal. [19]
Most importantly, nothing here contributes anything new to the
understanding of the pointer operators and the branching rates, or to
making a continuous-time description of the branching process itself.

The pointer operators for multi-particle systems employed here were
constructed to avoid having the anomalous branching induce any spurious
correlations on independent particles. In simple examples with the quantum
state of the system in the opposite limit, i.e. the zero-temperature
crystal,  the branching algorithm then also gave a reasonable results. I
have not provided an analysis of the behavior of partially correlated
particles. A proper description for collections of identical particles also
has not been given here.

In the artificial model constructed here to show that a no-collapse picture
can produce correct measure-proportional macroscopic probabilities, I have
built up pointer operators out of single-particle operators. That procedure
is easy, but it lacks a strong physical motivation and is very unlikely to
be suitable in general. A more developed theory obviously should not be
expressed in terms of individual particle states or any other decomposition
which becomes arbitrary in a general case. Perhaps a more suitable set of
pointer operators would depend on four-momentum density without reference
to constituents, along lines being pursued in collapse pictures. [8]
Construction of such a model is well beyond my capabilities.

The growth rates of the pointer operators ought to be expressed by some
operator which transforms with other physical rates if a relativistic
theory is to be obtained.  The indications of rest-mass dependence of the
collapse rate in collapse interpretations [8, 9, 10] suggest that the
growth rate $1/\tau$ for a given $P_\Gamma$ could be replaced by an
operator $\epsilon H/\hbar$ where $\epsilon$ is a small positive number and
H is understood to be the full Hamiltonian (i.e. whatever appears in the
gravitational source term, typically dominated by the rest-mass term) for
the particular subsystem on which $P_\Gamma$ acts. If the anomalous
branching rate for single particles is to fall within the rather broad
range of postulated rates in collapse theories [8, 19], we would need
$10^{-50} < \epsilon <10^{-40}.$

A non-uniform $\tau$ of this type would have some interesting, problematic
consequences. The branching rate would be higher on high-energy components
than on low-energy ones. Although that difference would in itself have no
direct effect on Tr($\rho$H) where $\rho$ is the global density matrix, it
would lead to an increase in the relative numbers of high-energy
sub-branches, giving a term $\epsilon <(\delta H)^2>_C/\hbar$  in the rate
of increase of $<H>_C$, where $<>_C$ again denotes averaging over
equal-weighted sub-branches. This effect is smaller than the direct energy
non-conservation predicted to arise from explicit collapse processes [9,
19] (an effect perhaps shared with the branching processes), and would lead
to no significant anomalies over many times the current lifetime of the
universe. For any collection of independent subsystems, the rate of energy
increase would come out to be the sum of the subsystem rates. Intriguing
problems would arise for the description of subsystems in a closed universe
whose net energy was identically zero.

\section{Conclusion: comparison with related ideas and observational
constraints}
The proposal is clearly mathematically and observationally distinct from
prior many-worlds pictures, which lack the prediction of anomalous loss of
interference. Ultimately, the parameters (e.g. $\epsilon$) fixing the rate
of anomalous interference loss should be measurable in mesoscopic
experiments if either the collapse approach or this approach is correct. I
shall argue that the non-linear many-worlds approach may have subtle
observational differences from the family of explicit collapse pictures
which share the prediction of anomalous decoherence.

The standard quantum probabilities here {\em emerge} under some limiting
conditions. There is a regime, illustrated in our toy examples, in which it
is meaningful to use outcome-counting to assign probabilities which are not
proportional to measure. However, the standard quantum mechanical
probabilities would result for observations on a time scale large compared
to the effective anomalous branching rate, so long as the commutators of
the pointer operators and the Hamiltonian are small enough, i.e so long as
condition (8) is met.

Given that as observers we are intrinsically limited to the size-time scale
on which experiments in the past have always given Born probabilities, the
most reasonable hope for finding different predictions would be to find
some experiment in which the commutators of the relevant pointer operators
and the Hamiltonian were large, so that condition (8) would be violated.
That would require some way of arranging for the state to maintain a rapid
decrease in the log of the maximum measure along any pointer sub-space
simply due to ordinary quantum dynamics. That decrease would have to be
unequal between different macro-outcomes and would have to be maintained
over the entire course of the experiment. That leaves (slightly) open both
the unpleasant possibility that in a more developed theory some case could
show probabilities already so far from known observations as to rule out
this approach and the pleasant possibility that some case might predict
probabilities subtly but measurably distinct not only from standard quantum
mechanics (which lacks the anomalous interference loss) but also from the
explicit collapse pictures.

It may seem that an approach which invokes non-linearity without getting
rid of many worlds has combined the worst features of two types of theory.
However, the acceptance of non-linearity seems to be a necessity if one is
to obtain the correct probabilities from the dynamical equations. The
choice between many-worlds and collapse is not dictated by any known
observation. By not insisting on unique outcomes, we avoid some of the
ingredients of the collapse models which seem most distinct from ordinary
quantum mechanics, partially compensating for the currently less-developed
state of the no-collapse approach. Non-quantized fields, hypotheses about
constructs outside the dynamical equations, and explicit stochastic
constituents are all avoided, as hoped for in early versions of many-worlds
pictures. [11] As Squires noted, [22] dropping the collapse hypothesis also
avoids the problem of requiring prior non-local correlations in random
collapse-generating fields. Any no-collapse picture, including this one,
avoids postulating unobserved state-reduction, although at the obvious cost
of postulating state-components unobserved by a given macroscopic observer.

Of course, if one or the other non-linear approach is found to flow in a
natural way from a deeper understanding of constituent physics (e.g.
gravity), no such mere postulates will be required. On the other hand, if a
full, consistent theory of all observed interactions were found to have the
form of a linear quantum field theory, then both approaches might be left
with the need for some highly arbitrary assumptions, without prospects of
confirmation from another line of reasoning.

Proposals which simply add the usual probability rule to a purely linear
dynamics cannot distinguish {\em in principle} whether a single-world [23]
or multi-world [16] interpretation is correct. Likewise, the Bohm
interpretation [24] is unable in principle to distinguish if there is one
``real" coordinate point guided by the wave or an ensemble of such points,
since the actual coordinates have no effect on the linear evolution of the
wave. In contrast, I have pointed to a route by which the non-standard
collapse and non-standard no-collapse pictures might conceivably be
distinguished.

The point has been made before that proposed experimental tests designed to
distinguish explicit collapse models from standard quantum mechanics would
determine only whether an additional source of interference loss exists,
not whether all but one of the resulting components of the state disappear.
[10]  This point is not a mere formality, because a non-linear no-collapse
account may give a straightforward mechanism for directly explaining
quantum probabilities in terms of simple numbers of non-interfering
outcomes arising directly from the dynamics, without postulating any
stochastic non-quantized fields.
\vskip24pt
\noindent{\Large\bf Acknowledgments}
I thank J. Ellis for a stimulating conversation, and an anonymous referee
for extraordinarily careful and thorough constructive criticism, which was
essential to getting this paper in respectable shape.
\vskip24pt
\noindent{\Large\bf References}

\begin{enumerate}
\item{W. H. Zurek, {\em Physics Today} {\bf 44}(1991) 36.}
\item{L. E. Ballentine, {\em Found. Phys.} {\bf 3} (1973) 229.}
\item{N. Graham, in The Many-Worlds Interpretation of Quantum
Mechanics(Princeton U., Princeton, 1973) pp. 229-553.}
\item{J. Ellis, N. E. Mavromatos, and D. V. Nanopoulos, {\em Mod. Phys. Lett.}
{\bf A10}(1995) 425.}
\item{J. Ellis, N. E. Mavromatos, and D. V. Nanopoulos, J. Chaos, {\em Solitons
and Fractals}, {\bf 10} (1999) 345.}
\item{P. Pearle, {\em Phys. Rev. A} {\bf 48}(1993) 913.}
\item{P. Pearle, in {\em Perspectives on Quantum Reality R. Clifton}, Ed.
(Kluwer, Dordrecht, 1996) pp. 93-109.}
\item{P. Pearle and E. Squires, {\em Found. Phys.} {\bf 26} (1996) 291.}
\item{P. Pearle, J. Ring, J. I. Collar, and F. T. Avignone, {\em Found. Phys.},
in press.}
\item{P. Pearle, in {\em Open Systems and Measurement in Relativistic Quantum
Mechanics} F. Petruccione, H. P. Breuer, Ed. (Springer Verlag), in press.}
\item{H. Everett, {\em Rev. Mod. Phys.} {\bf 29} (1957) 141.}
\item{B. DeWitt and N. Graham, {\em The Many-Worlds Interpretation of Quantum
Mechanics.}  (Princeton U., Princeton, 1973).}
\item{M. Gell-Mann and J. B. Hartle, {\em Phys. Rev. D} {\bf 47}(1993) 3345.}
\item{F. Dowker and A. Kent, {\em J. Stat. Phys.} {\bf 82}(1996) 1575.}
\item{W. H. Zurek, Phil. Trans. {\em R. Soc. Lond. A} {\bf 356} (1998) 1793.}
\item{S. Saunders, in {\em Perspectives on Quantum Reality} R. Clifton, Ed.
(Kluwer, Dordrecht, 1996) pp. 125-142.}
\item{D. Deutsch, {\em Int. J. Theor. Phys.} {\bf 24} (1985) 1.}
\item{D. Albert and B. Loewer, {\em Synthese} {\bf 77}(1988) 195.}
\item{G. C. Ghirardi, A. Rimini, and T. Weber, {\em Phys. Rev. D} {\bf 34}
(1986) 4701.}
\item{A. J. Leggett, {\em Found. Phys.} {\bf 18} (1988) 939.}
\item{J. Butterfield, G. N. Fleming, G. C. Ghirardi, and R. Grassi, {\em
Int. J.
Theor. Phys.} {\bf 32} (1993)  2287.}
\item{E. Squires, {\em The Mystery of the Quantum World.}  (Institute of
Physics,
Bristol, 1994).}
\item{N. D. Mermin, {\em Am. J. Phys.} {\bf 66} (1998) 753.}
\item{D. Bohm and B. J. Hiley, {\em The Undivided Universe}.  (Routledge,
London,
1993).}
\end{enumerate}

\end{document}